# First International Conference on Diffusion in Solids and Liquids

6-8 of July, 2005
Aveiro - Portugal

*Proceedings*

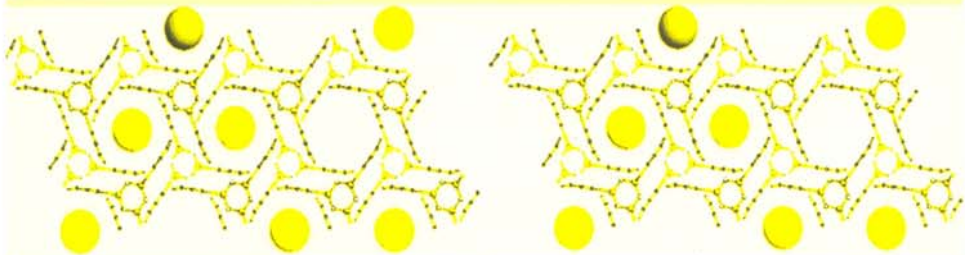

Editors:

**Andreas Öchsner**

**José Grácio**

**Frederic Barlat**

Proceedings of the
1st International Conference on Diffusion in Solids and Liquids
"DSL-2005"

Volume I

*Editors:*

Andreas Öchsner
José Grácio
Frédéric Barlat

Design – *Marcia Öchsner*

# PARAMETERS OF MICRODIFFUSION IN F.C.C.-Ni–Mo SOLID SOLUTIONS

Sergiy BOKOCH[1,*], Mykola KULISH[1], Valentyn TATARENKO[1,2], Taras RADCHENKO[2]

[1] Taras Shevchenko Kyyiv National University, 6 Academician Glushkov Ave., UA-03022 Kyyiv-22, Ukraine
[2] G.V. Kurdyumov Institute for Metal Physics, N.A.S.U., 36 Academician Vernadsky Blvd., UA-03680 Kyyiv-142, Ukraine

[*] Corresponding author. Fax: +380 44 424 25 61; Email: srg@univ.kiev.ua

**Abstract**

The time evolution of diffuse X-ray scattering intensities conditioned by the short-range order (SRO) in Ni–11.8 at.% Mo solid solutions is investigated. As shown, the transition from a quenched (nonequilibrium) state to the equilibrium one is accompanied by the complex time reorganization of various SRO types (namely, $N_2M_2$, $N_4M$, $N_2M$). Computer simulations of local atomic configurations in alloy with use of the Monte Carlo method, inhomogeneous-SRO model, and calculation of electron density of states (DOS) at the Fermi level give the opportunity of obtaining quantitative characteristics of atomic reconfigurations. The relaxation times of diffuse X-ray scattering intensities and Fourier components of atom jumps for different wave vectors **k** in reciprocal space are determined. Based on the microdiffusion model for f.c.c. solid solution, the micro- and macroscopic parameters of the atom migration of both components in Ni–Mo solid solutions are investigated.

*Keywords:* Basics of diffusion; Calculation of diffusion coefficients; Short-range order; X-ray diffuse scattering; Monte Carlo simulation

## 1 Introduction

Kinetics of relaxation of atoms' configurations into the equilibrium state of alloys is usually considered as providing by the macroscopic atom migration onto significant distances with predetermined diffusion coefficients. Microscopic diffusion parameters concerned with probabilities of elementary atom jumps for various components on sites of a regular lattice in alloy are insufficiently investigated. The special emphasis is given to microscopic models, which enable to determine the mechanisms of diffusion realized during a short-range order (SRO) relaxation in an arrangement of atoms in alloys. Peculiarity of this SRO relaxation in alloys is provided that relaxation is realized by the atom jumps onto distances, which are commensurable with intersite distances. Activation of such a migration dispenses the high temperatures.

During many years, the substitutional f.c.c-Ni–Mo alloys of stoichiometry composition attract the attention of many researchers because of a whole series of salient features in their ordering. Independently on components' concentration in Ni–Mo alloys at the initial stages of formation of the ordered structures after quenching, there is inconsistency between the $\{1\frac{1}{2}0\}$-type short-range order and diffraction patterns of equilibrium long-range ordered (LRO) structures [1–3]. Despite the long-term investigations of the mentioned phenomena, the mechanisms of short-range order transformations are not determined. As result of the complex character of SRO evolution in Ni–Mo solid solutions, it is important to study the mechanisms of SRO relaxation that is possible only at the quantitative determination of microdiffusion parameters.

One of the methods of studying such diffusion parameters for alloys is the method of investigation of isothermal dependences of residual electroresistance [4]. However, it is necessary to take into account that this method is limited in the analysis of dependence of relaxation times of electroresistance on separate wave vectors **k**, which characterize the short-range order. Relaxation times of a residual electroresistance are integrated characteristics of electronic-waves' scattering on a whole set of ordering wave-vector stars $\{\mathbf{k_S}\}$. In details, the relaxation of separate short-range order types can be investigated only by means of studying kinetics of diffuse radiation scattering intensities (for X-rays or thermal neutrons).

## 2 Experimental procedures

Kinetics of diffuse X-ray scattering is studied for a single-crystal Ni–11.8 at.% Mo specimen (of $2.5\times6\times6$ mm$^3$ in size) grown in induction furnace in argon atmosphere within the alundum crucible with a conic seed region by Czochralski method. The surface of a single crystal was cut out along a high-symmetry direction, and its external side coincided with a (001)* plane. Initially, the specimen was annealed at the temperature ($T_q \approx 1223$ K), which is higher than the temperature of order–disorder transformation, during 2 hours and after it was quenched into 10% NaOH solution in a water at the room temperature. Then, a quenched specimen was heated up to various annealing temperatures ($T_a \in 323$–$423$ K) and was annealed during a certain time, and it was again quenched.

Diffuse X-ray scattering was measured with use of the hard Mo$K_\alpha$ radiation. The separation of different components conditioned by the SRO, static and dynamic displacements, which modulate the diffuse scattering in alloys, are carried out with use of the Cohen–Georgopoulos method [5].

Diffuse X-ray scattering intensity distributions were studied in the positions, which are equivalent to $\frac{1}{4}(420)$ (Ni$_2$Mo$_2$), $\frac{1}{5}(420)$ (Ni$_4$Mo), $\frac{1}{3}(420)$ (Ni$_2$Mo) points, and also within their vicinities close to the (001)* plane of the first Brillouin zone in a reciprocal space (see Fig. 1).

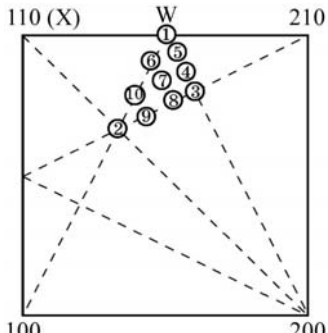

Figure 1: Positions of points at issue in the 1$^{st}$ Brillouin zone of f.c.c.-lattice ($1$—$\frac{1}{4}(420)$; $2$—$\frac{1}{3}(420)$; $3$—$\frac{1}{5}(420)$).

Based on the obtained diffuse intensities' distributions, the Warren–Cowley SRO parameters, $\alpha(lmn) \equiv \alpha(\mathbf{r})$, are calculated as follows:

$$\alpha(lmn) = K(\mathbf{r}) \sum_{\mathbf{k} \in BZ} I_{\text{SRO}}(\mathbf{k}) \cos(2\pi \mathbf{k} \cdot \mathbf{r}), \quad (1)$$

where $I_{\text{SRO}}(\mathbf{k}) \equiv I_{\text{SRO}}(h_1, h_2, h_3)$—diffuse scattering intensity concerned with SRO, $lmn$—co-ordinates of f.c.c. lattice sites, $K$—normalizing factor. SRO parameters, $\alpha(lmn)$, are concerned with the occupation probability for the $A(B)$-type atom, $P_{lmn}^{AB}$, on a distance $\mathbf{r}(lmn)$ on condition that 'zero' site is occupied by the $B(A)$ atom: $\alpha(lmn) = 1 - P_{lmn}^{AB}/c_B$, where $c_B$—relative concentration of $B$ atoms.

## 3 Kinetics of diffuse X-ray scattering intensity distribution

In Figure 2, the diffuse scattering intensity distribution conditioned by SRO for Ni–11.8 at.% Mo alloy are shown for two states. As visible, in a quenched state, intensity $I_{\text{SRO}}(h_1, h_2, 0)$ is mainly localized in $\{1\frac{1}{2}0\}$ positions and is large in other points. In superstructural points $\frac{1}{3}(420)$ and $\frac{1}{5}(420)$, diffuse scattering intensity is less. So, the obtained value of $I_{\text{SRO}}(h_1, h_2, 0)$ specifies that, in a quenched state, the SRO type corresponds to the coexistence of Ni$_2$Mo$_2$, Ni$_4$Mo, and Ni$_2$Mo structures.

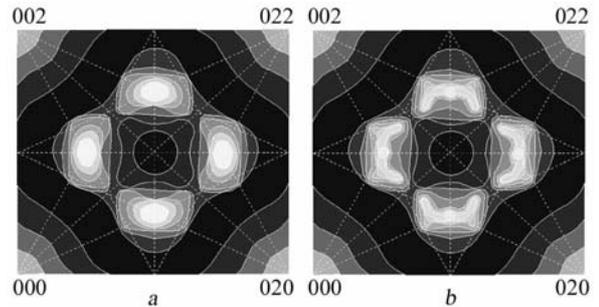

Figure 2: SRO intensity distributions, $I_{\text{SRO}}(h_1, h_2, 0)$, for various states of Ni–11.8 at.% Mo alloy: quenched from $T_q = 1223$ K (*a*) or quenched from $T_q = 1223$ K and then annealed at $T_a = 373$ K during 60 minutes (*b*).

In contrast to the quenched state (Fig. 2*a*), in an annealed state (Fig. 2*b*), intensity $I_{\text{SRO}}(h_1, h_2, 0)$ has maxima in $\{1\frac{1}{2}0\}$ and $\frac{1}{5}(420)$ positions and is smaller in $\frac{1}{3}(420)$ point. In Figure 3, kinetics of the Warren–Cowley SRO parameters, $\alpha_{lmn}(t)$, is shown for five co-ordination shells that has been calculated by distributions of $I_{\text{SRO}}(h_1, h_2, 0)$ (Fig. 2).

It is necessary to pay into account that SRO parameters $\alpha(lmn)$ for all co-ordination shells behave nonmonotonely. One can see that transition from the quenched (nonequilibrium) SRO state to equilibrium one is accompanied by the alloy disordering that is characterized by the decreasing of $\alpha(lmn)$ values on all co-ordination shells within the certain annealing-time interval. Comparison of obtained SRO parameters, $\alpha(lmn)$ (Fig. 3), with SRO parameters for the LRO $D1_a$, $D0_{22}$, $N_3M$, $N_2M_2$ structures testifies to presence mainly $D1_a$-type SRO in an equilibrium state (Fig. 2*b*).

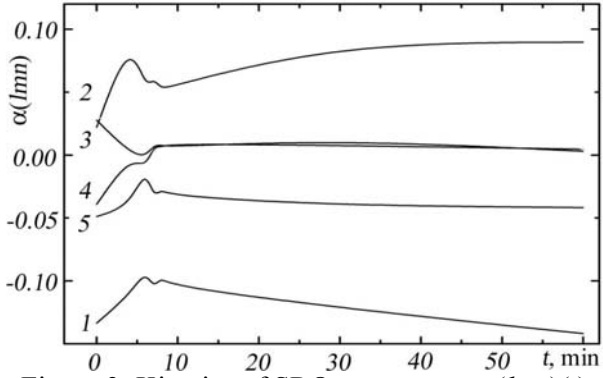

Figure 3: Kinetics of SRO parameters $\alpha_i(lmn)(t)$ for n-th shell (n = 1–5) in Ni–11.8 at.% Mo alloy quenched from 1223 K and annealed at 373 K.

Hence, the time evolution of diffuse X-ray scattering distributions manifests complex kinetic behaviour of SRO parameters. Such a behaviour of $\alpha(\mathbf{r})(t)$, depending on annealing time $t$, corresponds to presence the static concentration waves with different wave vectors $\mathbf{k}$ ($\langle 1\tfrac{1}{2}0 \rangle$, $\tfrac{1}{5}\langle 420 \rangle$, and $\tfrac{1}{3}\langle 420 \rangle$) during the atom ordering in alloy.

## 4 Computer simulation of local atomic configurations in alloy

It is evident that the concurrent presence of several SRO types in an alloy and their complex time evolution must be characterized by the complicated kinetic pattern of local atomic configurations. Quantitative calculations of reorganization of atomic configurations have been carried out with use of both Monte Carlo method simulation [6] and inhomogeneous-SRO model.

The Monte Carlo method is based on the use of experimental values of SRO parameters, $\alpha(lmn)$ Fig. 3. In addition, the modelled f.c.c.-lattice array is consisted of $5\times10^5$ sites. The modelling alloy with a random distribution of atoms is generated as initial state. For subsequent stages, atoms of different kinds are randomly chosen and exchanged by places, and the following parameter is calculated:

$$R = \sum_{lmn} (\alpha_{lmn}^{calc} - \alpha_{lmn}^{exp})^2 \Big/ \sum_{lmn} (\alpha_{lmn}^{exp})^2, \quad (2)$$

where $\alpha_{lmn}^{calc}$ are the Warren–Cowley SRO parameters calculated after rearrangement, $\alpha_{lmn}^{exp}$ — experimental set of SRO parameters (Fig. 3). Duration of Monte Carlo procedure is limited to the value of $R \approx 10^{-6}$. In Figure 4, results of Monte Carlo simulation of a local atomic structure in Ni–11.8 at.% Mo alloy for two SRO states are presented. As shown in Fig. 4, during a process of the SRO relaxation from the quenched state to equilibrium one at 373 K, its structure can be presented as a cluster mixture of $Ni_2Mo$, $Ni_3Mo$, and $Ni_4Mo$ types. Such SRO structure is similar to the structure, which is existing at initial ordering stages of $Ni_3Mo$ alloys [2].

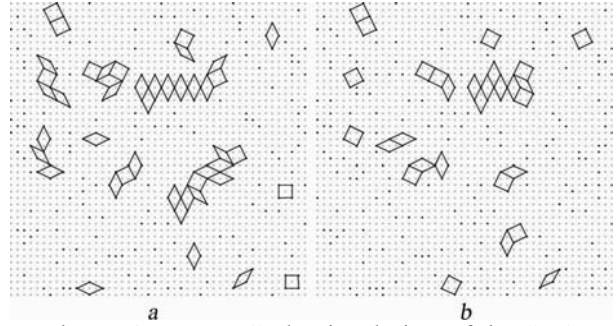

Figure 4: Monte Carlo simulation of the SRO structure in Ni–11.8 at.% Mo alloy for two states: *a*—quenched from 1223 K, *b*—annealed at 373 K during 60 minutes (○, ●—Ni and Mo atoms, respectively; 'rhombuses'—$D0_{22}$ and $Ni_2Mo$ clusters, 'squares'—$Ni_4Mo$-type clusters).

The cluster structure of local atomic configurations in Ni–Mo alloys presupposes their consideration within the framework of the inhomogeneous-SRO model. Thus, it is supposed that the alloy consists of disordered matrix with SRO parameters $\alpha_{lmn}^{(0)}$, where ordered *j*-type clusters with parameters $\alpha_{lmn}^{(j)}$ are injected ($j = 1, 2, \ldots$). If the concentration of *j*-type clusters is equal to $\beta^j$, it is possible to write the following relationships:

$$\sum_j \beta^j = 1, \quad x = \sum_j \beta^j x_j, \quad y = \sum_j \beta^j y_j,$$
$$y\alpha_{lmn} = \sum_j \beta^j y_j \alpha_{lmn}^j, \quad x\alpha_{lmn} = \sum_j \beta^j x_j \alpha_{lmn}^j, \quad (3)$$

where $x, y$ and $x_j, y_j$ are the relative concentrations of both components in alloy and in *j*-type cluster, respectively. Neglecting underdeveloped ordering of a matrix ($\alpha_{lmn}^{(0)} \cong 0$), for determination of the concentration of ordered clusters, it is necessary to calculate the following square-law functional:

$$V = \sum_{\mathbf{r}} (x\alpha_{\mathbf{r}} - \sum_j x_j \alpha_{\mathbf{r}}^{(j)} \beta^{(j)})^2 + \\ + \sum_{\mathbf{r}} (y\alpha_{\mathbf{r}} - \sum_j y_j \alpha_{\mathbf{r}}^{(j)} \beta^{(j)})^2 \quad (\mathbf{r} = \mathbf{r}(l,m,n)). \quad (4)$$

Under the condition that the functional $V$ has a minimum against the concentration of clusters,

$\beta^{(j)}$, — $\partial V/\partial \beta^j = 0$, it is possible to obtain the set of equations, which allows to determine the concentration of clusters of different order types, $\beta^{(j)}$. Results of such a calculation are shown in Fig. 5 for Ni–11.8 at.% Mo alloy quenched from 1223 K and annealed at 373 K.

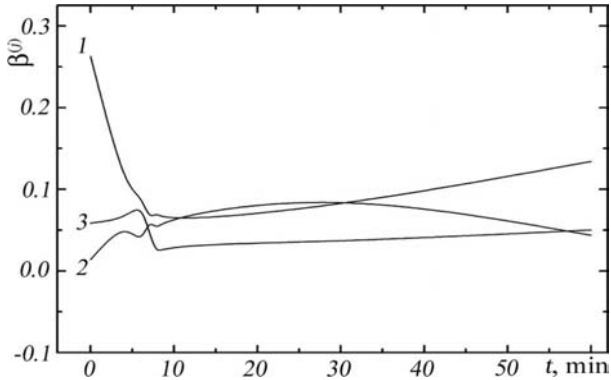

Figure 5: Time relaxation of the cluster concentrations, $\beta^{(j)}$, for $Ni_4Mo$ (*1*), $Ni_3Mo$ (*2*), $Ni_2Mo$ (*3*) clusters in Ni–11.8 at.% Mo alloy quenched from 1223 and annealed at 373 K.

As visible from Fig. 5, at the earliest stage of SRO relaxation, the sharp decreasing of the content of $Ni_4Mo$-type clusters is observed. In subsequent stages, the concentration $\beta(Ni_4Mo)$ is smoothly increasing during a long time, and near the equilibrium state, it increases essentially. Concentration of clusters of two other types, $\beta(Ni_3Mo)$ and $\beta(Ni_2Mo)$, is insignificant (~5–10%) during a relaxation. The basic feature of a concentration relaxation of different-type clusters, $\beta^j$, is their decreasing within the certain annealing-time interval that testifies to the disordering of alloy during the SRO evolution into the equilibrium state. This result agrees with kinetics of SRO parameters, $\alpha(lmn)$ (see Fig. 3).

## 5 Electron density of states

As known, the electron density of states (DOS) for alloys is in accordance with their phase-stability conditions [3, 4]. Calculation of DOS, $g(E)$, for Ni–11.8 at.% Mo alloy is based on the method of cluster expansion of the Green function of disordered systems [4], which uses the experimental values of SRO parameters $\alpha(lmn)$ (Fig. 3).

In Figure 6, the calculated DOS, $g(E)$, are presented for two states of Ni–11.8 at.% Mo alloy. As shown, in annealed state (Fig. 6*b*), the DOS at Fermi level, $E_F$, is less than in a quenched state that testifies to the stability of mixture of different order-type clusters in equilibrium (see Sec. 4).

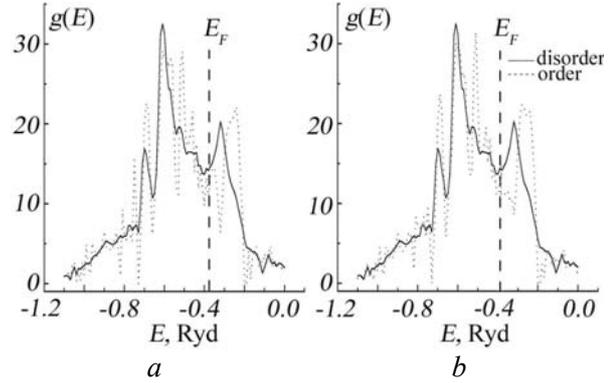

*a*  *b*

Figure 6: Calculated DOS for Ni–11.8 at.% Mo alloy for two states: *a*—quenched from 1223 K and *b*—annealed at 373 K during 60 minutes. (Continuous line presents the DOS for disordered state, and dashed line corresponds to the ordered state.)

## 6 Microdiffusion parameters

### 6.1 Microdiffusion parameters within the Fourier representation

As known [7], the presence of SRO in alloys, where any-type atom located at zero site, predetermines the concentration inhomogeneity on all co-ordination shells. Unlike the continuum concentration gradients, the sizes of such concentration inhomogeneities during the diffusion SRO relaxation is characterized by both the microscopic atom mobility and probabilities of atom jumps onto intersite distances. For obtaining microscopic characteristics of such a diffusion, investigation of the time dependence of two-particle correlation function (SRO parameters) arrangement of atoms is necessary. It is possible due to the investigation of diffuse X-ray scattering intensity relaxation conditioned by SRO, $I_{SRO}(\mathbf{k},t)$ [7].

As known (see [7] and Refs. therein), in general for a binary alloy (with any mobilities of components), an expression for the time dependence of SRO intensity, $I_{SRO}(\mathbf{k},t)$, can be written as follows:

$$I_{SRO}(\mathbf{k},t) = I_{eq}(\mathbf{k},\infty) + A_1 \exp[-t/\tau_1(\mathbf{k})] + \\ + A_2 \exp[-t/\tau_2(\mathbf{k})] + A_3 \exp[-t/\tau_3(\mathbf{k})], \quad (5)$$

where $I_{SRO}(\mathbf{k},t)$ and $I_{eq}(\mathbf{k},\infty)$—instantaneous (non-equilibrium) and equilibrium values of diffuse scattering intensities, respectively; $A_1$, $A_2$, $A_3$— 'weights' of corresponding relaxation processes; $\tau_1(\mathbf{k})$, $\tau_2(\mathbf{k})$, $\tau_3(\mathbf{k})$—their relaxation times (only two ones are independent with each other), which are related to the Fourier components ($\lambda_1(\mathbf{k})$ and $\lambda_2(\mathbf{k})$) of probabilities of atom jumps of components (per unit of time) in a binary alloy as follows:

$$\tau_1(\mathbf{k}) \approx [2\lambda_1(\mathbf{k})]^{-1}, \quad \tau_2(\mathbf{k}) \approx [2\lambda_2(\mathbf{k})]^{-1},$$
$$\tau_3(\mathbf{k}) \approx [\lambda_1(\mathbf{k}) + \lambda_2(\mathbf{k})]^{-1}. \quad (6)$$

In case of almost identical mobility of both components ($\tau_1(\mathbf{k}) \approx \tau_2(\mathbf{k})$), an expression (5) will be following:

$$I_{SRO}(\mathbf{k},t) = I_{eq}(\mathbf{k},\infty) + A_1 \exp[-t/\tau_1(\mathbf{k})] + A_2 \exp[-t/\tau_2(\mathbf{k})], \quad (7)$$

and in case of realization of the condition $\tau_1(\mathbf{k}) \gg \tau_2(\mathbf{k})$, we have:

$$I_{SRO}(\mathbf{k},t) = I_{eq}(\mathbf{k},\infty) + A_1 \exp[-t/\tau_1(\mathbf{k})]. \quad (8)$$

In Figure 7, the time evolution of SRO intensity, $I_{SRO}(\mathbf{k}, t)$, is shown for a high-symmetry point in the first Brillouin zone W(1 ½ 0) for Ni–11.8 at.% Mo alloy quenched from 1223 K and annealed at 373 K.

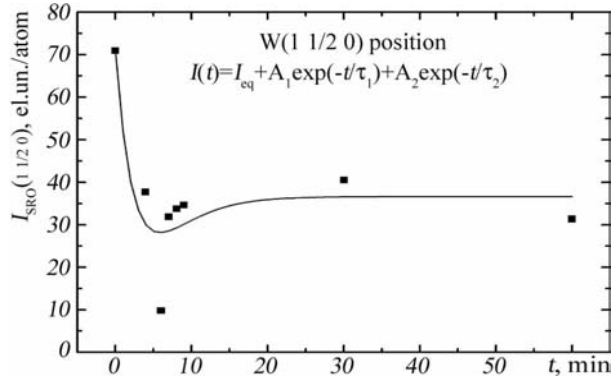

Figure 7: Time evolution of the SRO intensity $I_{SRO}(\mathbf{k}, t)$ for a point W(1 ½ 0).

Describing experimental values of diffuse intensity, $I_{SRO}(\mathbf{k}, t)$, in a vicinity of points at issue within the approximation (7) and using expression (6), the Fourier components of atom-jump probabilities per unit of time, $\lambda_{Ni}(\mathbf{k})$ and $\lambda_{Mo}(\mathbf{k})$, are obtained. For example, for high-symmetry point W(1 ½ 0), $\lambda_{Ni}(W) = 2.49 \times 10^3$ sec$^{-1}$ and $\lambda_{Mo}(W) = 4.80 \times 10^3$ sec$^{-1}$. The obtained values of $\lambda_{Ni}(\mathbf{k})$ and $\lambda_{Mo}(\mathbf{k})$ are non-monotonely increasing with approaching to the first Brillouin zone that testifies to the action of non-vacancy mechanism of diffusion during the SRO relaxation of Ni–Mo solid solutions at the annealing temperatures at issue.

### 6.2 Fourier originals of microdiffusion parameters

Taking into account the complexity of processes of SRO transformation at low annealing temperatures in Ni–Mo alloys (see sections 3–5 and subsection 6.1), there is necessity to involve the correct diffusion model, which assumes the studying parameters of microdiffusion. This model must take into account the opportunity of commensurable mobility of both components in alloy as well as the opportunity of atom jumps within the several co-ordination shells.

In a given work, the potential field, $\psi_S(\mathbf{r}_n)$, conditioned by the point microinhomogeneity of impurity atom in zero site, is supposed and distributed up to 6 co-ordination shells; probabilities of elementary atom jumps into the site $\mathbf{r}$ from all other sites $\{\mathbf{r}_n\}$, $-\Lambda_S^0(\mathbf{r} - \mathbf{r}_n)$, are non-vanishing for distances $\{|\mathbf{r} - \mathbf{r}_n|\}$, which do not exceed the second nearest-neighbouring shell radius:

$$\Lambda_S(\mathbf{r} - \mathbf{r}_{I,II}) \neq 0, \quad \psi_S(\mathbf{r}_{n=1,\ldots,6}) \neq 0 \quad (s=Ni, Mo). \quad (9)$$

Making an inverse Fourier transformation of $\lambda_{Ni,Mo}(\mathbf{k})$ obtained according to (6), probabilities of the atom jumps, $-\Lambda_S(\mathbf{r}_n)$ (in a nonideal solid solution with a SRO), up to 8 co-ordination shells are determined. According to Ref. [7], it is possible to write as follows:

$$\Lambda_S(\mathbf{r}) \cong \sum_{\mathbf{r}_n} \Lambda_S^0(\mathbf{r} - \mathbf{r}_n)(c_B(1-c_B)/k_B T)\psi_S(\mathbf{r}_n), \quad (10)$$

where $-\Lambda_S^0(\mathbf{r} - \mathbf{r}_n)$—the probability of an elementary jump of s-kind atom from a site $\mathbf{r}_n$ into the site $\mathbf{r}$ in case of the ideal solid solution ('without SRO'); $\psi_S(\mathbf{0}) \equiv k_B T/c_B(1-c_B)$ is the Fourier original of microinhomogeneity potential at the zero site in ideal solid solution with absolute temperature $T$; $k_B$—Boltzmann constant.

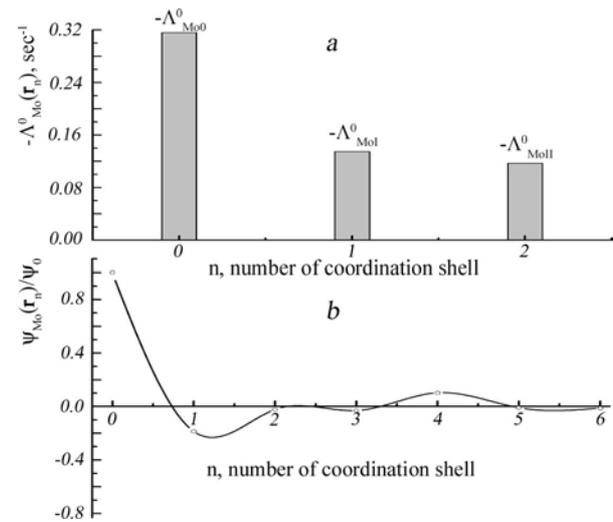

Figure 8: Probabilities of Mo atom jumps in ideal solid solution $\{-\Lambda_{Mo}^0(\mathbf{r}_n)\}$ (*a*) and the normalized potential function $\psi_{Mo}(\mathbf{r}_n)/\psi_{Mo}(\mathbf{0})$ (*b*).

Solving set of nonlinear equations (10) (see Ref. [8]), the probabilities of atom jumps, $\{-\Lambda_S^0(\mathbf{r}_n)\}$, and the normalized potential functions, $\{\psi_S(\mathbf{r}_n)/\psi_S(\mathbf{0})\}$, for atoms of both components are obtained. For instance in Fig. 8, the mentioned dependences for Mo atoms are presented.

As shown in Fig. 8*a*, the probabilities of Mo-atom jumps into the first and second co-ordination shells ($-\Lambda_{\text{MoI}}^0$ and $-\Lambda_{\text{MoII}}^0$) are commensurable on magnitude that testifies to the nothingness of the vacancy mechanism of diffusion at annealing temperatures close to the room temperature. Potential function (Fig. 8*b*) behaves non-monotonely, and on $5^{\text{th}}$ and $6^{\text{th}}$ co-ordination shells, it can be neglected.

Accordingly [7], there are relationships between the microscopic probabilities of atom jumps in ideal or nonideal solid solutions and macroscopic (continuum) diffusion ($D$) or self-diffusions ($D^*$) coefficients, respectively:

$$D_S \approx \sum_{n=I}^{\infty} -\frac{\Lambda_S(\mathbf{r}_n)\mathbf{r}_n^2 z_n}{6}, \quad D_S^* \approx \sum_{n=I}^{\infty} -\frac{\Lambda_S^0(\mathbf{r}_n)\mathbf{r}_n^2 z_n}{6}, \quad (11)$$

where $\mathbf{r}_n$ and $z_n$ are the radius-vector and co-ordination number n-th shell, respectively. Substituting the magnitudes of $\Lambda_S(\mathbf{r}_n)$ and $\Lambda_S^0(\mathbf{r}_n)$ (obtained for three annealing temperatures 323, 373 and 423 K for Ni–Mo solid solutions) in (11), the diffusion ($D$) and self-diffusion ($D^*$) coefficients as well as their activation energies are calculated. Values for Ni atoms are presented in Table 1.

Table 1: Diffusion parameters of Ni atoms in Ni–11.8 at.% Mo solid solution.

| $T_a$, K | $D$, sm$^2$/sec | $D^*$, sm$^2$/sec | $Q$, eV | $Q^*$, eV |
|---|---|---|---|---|
| 323 | $9.48 \times 10^{-16}$ | $7.08 \times 10^{-18}$ | 0.267 | 0.474 |
| 373 | $1.52 \times 10^{-15}$ | $3.12 \times 10^{-17}$ | | |
| 423 | $7.99 \times 10^{-15}$ | $3.48 \times 10^{-16}$ | | |

One can see that the values of activation energies of diffusion are small. It specifies the discrepancy between microdiffusion processes at the temperatures close to the room temperature and the vacancy-mechanism diffusion, which is observed at higher temperatures.

**Conclusion**

1. Kinetics of diffuse X-ray scattering intensity distribution for Ni–Mo solid solutions indicates that the SRO relaxation to the equilibrium state is accompanied by the transformation of its initial type to the $D1_a$-type SRO, *i.e.* to the order, which is observed in an alloy of stoichiometric composition Ni$_4$Mo [1, 3]. This conclusion agrees with results of computer simulation of local atomic ordering with use of both the Monte Carlo method and the inhomogeneous-SRO model. Thus, it is also obvious that the concentration of Ni$_4$Mo clusters is increasing with time. For such states close to the equilibrium, there is a minimum of electron DOS at the Fermi level that testifies to the stability of $D1_a$-type SRO in equilibrium state.

2. A SRO relaxation in Ni–Mo solid solutions within the temperature interval 323–423 K, it is not allowed to restrict the description by one relaxation time only; such a consideration demands using approximate two- or three-exponent models of relaxation of diffuse scattering intensity.

3. Complex changes of the Fourier components of probabilities of atom jumps, $\lambda_{\text{Ni}}(\mathbf{k})$ and $\lambda_{\text{Mo}}(\mathbf{k})$, in a vicinity of the $1^{\text{st}}$ Brillouin zone boundary and magnitudes of the values of macrodiffusion parameters obtained from $\lambda_{\text{Ni}}(\mathbf{k})$ and $\lambda_{\text{Mo}}(\mathbf{k})$ ($D$, $D^*$, $Q$, $Q^*$) is evidence of the SRO-relaxation mechanism at the temperatures close to the room temperature, which is not vacancy mechanism observed at higher temperatures.